\documentclass{aastex6}

\usepackage{graphicx}
\usepackage{txfonts}
\newcommand{\change}[1]{{#1}}
\newcommand{\changeb}[1]{{\bf #1}}
\begin{document}

   \title{Simulations of Alfv\'{e}n and kink wave driving of the solar chromosphere - efficient heating and spicule launching.}

   \author{C. S. Brady}
   \email{c.s.brady@warwick.ac.uk}    
   \author{T. D. Arber}

   \affil{Centre for Fusion, Space and Astrophysics
	University of Warwick, Coventry, CV4 7AL, UK\\
             }         

   \date{Submitted \today}
   \begin{abstract}
      Two of the central problems in our understanding of the solar chromosphere are how the upper chromosphere is heated and
      what drives spicules. Estimates of the required chromospheric heating, based on radiative and conductive losses 
      suggest a rate of ${\sim} 0.1 \mathrm{\:erg\:cm^{-3}\:s^{-1}}$ in the lower chromosphere dropping to  
      ${\sim} 10^{-3} \mathrm{\:erg\:cm^{-3}\:s^{-1}}$ in the upper chromosphere (\citet{Avrett1981}). The chromosphere is
      also permeated by spicules, higher density plasma
      from the lower atmosphere propelled upwards at speeds of ${\sim} 10-20 \mathrm{\:km\:s^{-1}}$, for so called
      Type-I spicules (\citet{Pereira2012,Zhang2012}), reaching heights of 
      ${\sim} 3000-5000 \mathrm{\:km}$ above the photosphere. A clearer understanding of chromospheric dynamics, its 
      heating and the formation of spicules, is thus of central importance to solar atmospheric science. For over
      thirty years it has been proposed that photospheric driving of MHD waves may be responsible for both heating
      and spicule formation. This paper presents results from a high-resolution MHD treatment of photospheric
      driven Alfv\'en and kink waves propagating upwards into an expanding flux tube embedded in a model chromospheric atmosphere.
      We show that the ponderomotive coupling from Alfv\'{e}n and kink waves into slow modes generates shocks which both heat 
      the upper chromosphere and drive spicules. These simulations show that wave driving of the solar chromosphere
      can give a local heating rate which matches observations and drive spicules consistent with Type-I observations
      all within a single coherent model.
   \end{abstract}
      \maketitle
\section{Introduction}

\change{The mechanisms by which the solar chromosphere are heated are still the subject of active debate. Alfv\'{e}n waves 
have been proposed as a possible heating mechanism \citep{Osterbrock1961,Ballegooijen2011} and observations show that there 
is enough Alfv\'{e}n wave energy generated in the convection zone \citep{Depontieu2007} to heat the chromosphere. Previous work
has examined this problem numerically. The coupling of circularly polarized Alfv\'{e}n waves to 
magneto-acoustic modes and shocks has been studied since the 1970's by Hollweg \citep{Hollweg1978,Hollweg1981,Hollweg1982}, 
and this was later extended to include white noise drivers in 1.5D \citep{Kudoh1999} or 2.5D \citep{Matsumoto2012}, \citep{Matsumoto2014}. While 
these simulations do address chromospheric heating, they were mainly concerned with the upper atmosphere and solar wind and 
so they generally do not present detailed results for chromospheric heating.} \citet{Arber2016} showed from 1D MHD simulations 
that chromospheric heating from Alfv\'{e}n waves is primarily due to shock 
heating from slow mode shocks generated by ponderomotive coupling.
\change{The ponderomotive force here is defined as the non-linear force generated by the gradient in the magnetic component
of MHD wave energy. Here the ponderomotive force density $F_{pm}$ is defined by $F_{pm} = -\nabla(B_\perp^2)/\mu_0$, where $B_\perp$ 
is the perturbation to the background magnetic field perpendicular to the equilibrium field \citet{Verwichte1999}.}

There is a comparable history of numerical work on the possible ways in which spicules can be launched. \citet{Haerendal1992} posited that they 
could be launched by ion-neutral collision effects, although the work of \citet{James2002} showed from 1D MHD simulations that this is unlikely. 
\citet{Depontieu2004} simulated the launching of spicules from the non-linear interactions of acoustic waves with a 2D solar atmosphere using a 
reduced model, finding that it is possible to explain spicule launching by this mechanism. \citet{Murawski2010} and \citet{Murawski2011} found from 
full 2D simulations that impulsive acoustic driving can lead to the formation of spicules. \citet{Cranmer2015} studied 1D simulations of spicule 
launching due to ponderomotive steepening of Alfv\'{e}n waves, also using a reduced model, finding that this mechanism can also produce realistic 
spicules.

Full 3D radiation-hydrodynamic simulations of the chromosphere, such as those using the BiFROST code \citep{Carlsson2016}, have shown the presence 
of shocks in the chromospheric cavity. Also recent
simulations of the whole chromosphere, transition region, corona and solar wind \citep{Matsumoto2012} demonstrate the importance of shocks below 
the transition region. However the complexity of these simulations has so far prevented a clear understanding of the underlying process as to how 
these shocks are formed and their precise role in matching the chromospheric heating requirements as a function of height. In this paper we 
present high-resolution simulations demonstrating that for a range of physically plausible parameters one can simultaneously 
produce chromospheric heating 
profiles and launch Type-I spicules with the correct length-scales, rise velocities and transverse oscillations. For each of these
(heating, spicule formation and oscillation) we compare the results with observations demonstrating that MHD wave driving of the solar chromosphere 
can self-consistently and accurately match these observations.

\section{Method}

\begin{figure}
\includegraphics[scale=0.3]{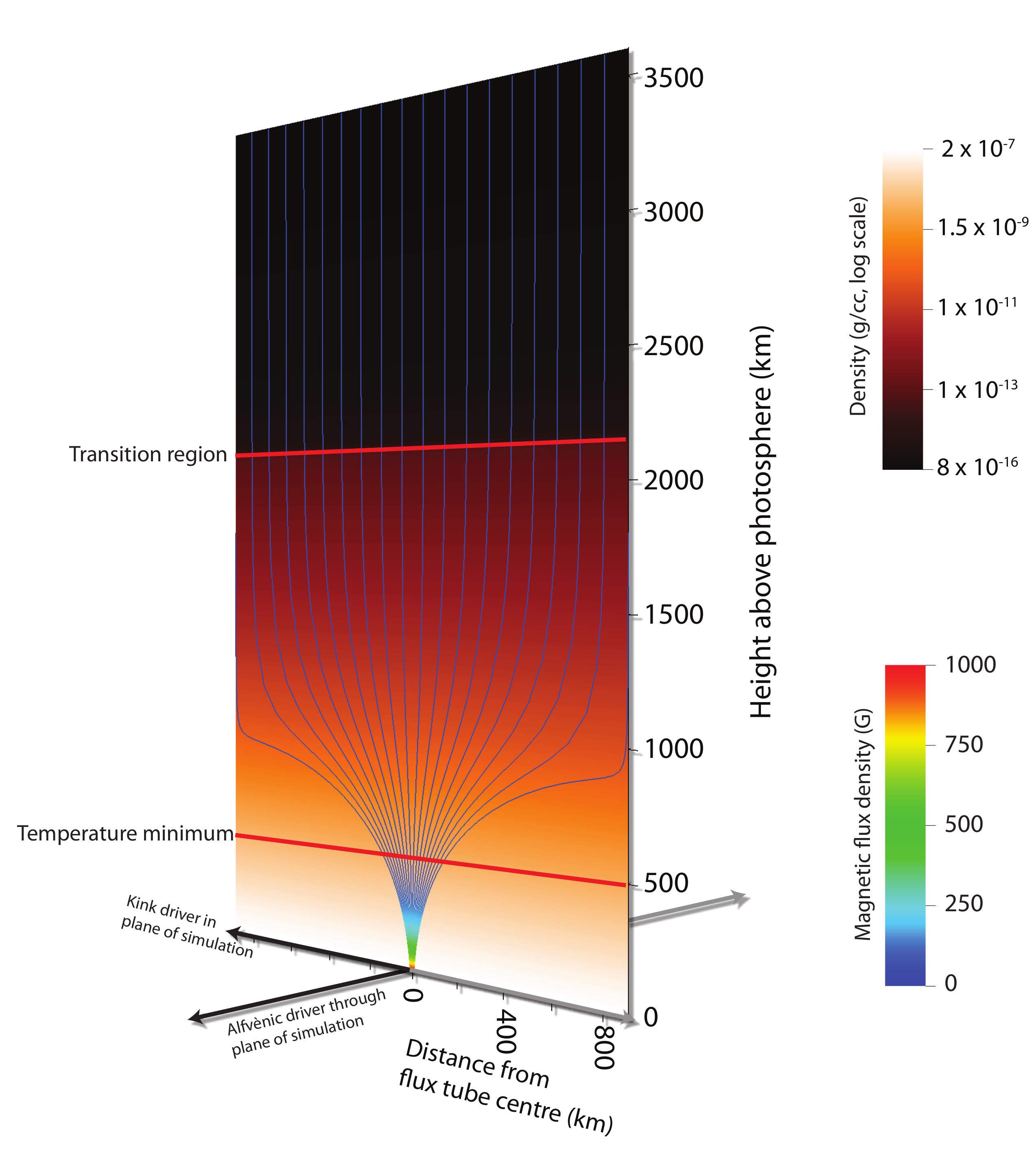}
\caption{Magnetic field configuration and strength in the initial conditions (coloured field lines). Also shown
is initial density (orange coloured plot). 
The direction of the two drivers used in the simulations is highlighted on the bottom boundary.}
\label{fig:tube}
\end{figure}

The aim of this paper is to evaluate the heating rate of the solar chromosphere combined with the properties of any spicules 
launched into the corona associated with the heating.  This is achieved by simulating the interaction of a spectrum of MHD waves 
propagating up into a 2.5D solar atmosphere model including partial ionisation, stratification and flux tube expansion. The code used is 
{\it Lare2d} \citep{Arber2001}. A fluid equilibrium is constructed \change{by setting up the Avrett and Loeser C7 temperature profile 
\citep{Avrett2008} and then integrating the density from the bottom boundary so that the atmosphere is in hydrostatic equilibrium. 
The domain extends 9Mm above the photosphere with a transverse width of 1.8 Mm and the results presented in this paper use a resolution of
4096 cells in the vertical ($y$) direction and 2048 horizontally (the $x$ direction). The position the centre of the flux patch is at $x=0$ 
and $y=0$, the base of the model photosphere, is 500 km below the temperature minimum. Throughout this paper 'height above the photosphere'
refers to height above $y=0$.
Convergence is tested by repeating a sample of simulations with 
double the resolution showing that results presented here are accurate to within 1\% on doubling the resolution.}
The simulation is run for a time of 1000 seconds, approximately 15 Alfv\'en transit times across the chromosphere, at which 
point it is found that the heating rate is converged.

\change{The ionisation state is calculated from a two level Athay potential model (\citet{Leake2005}, \citet{Thomas1961}) with the ionisation 
state and density being iterated until the atmosphere is in hydrostatic equilibrium.} A potential magnetic field is constructed numerically 
by solving Laplace's equation ($\nabla^2 \phi = 0$) for $\phi$, the magnetic scalar potential, subject to boundary conditions of 
transverse periodicity and 
a 10km flux patch on the photosphere resulting from setting $\phi=\phi_0 \exp(-x^2/\sigma^2)$ on the lower boundary
with $\sigma=10$ km (see figure \ref{fig:tube}). The magnetic field is then normalised to give a 
1kG peak field at the 
photosphere. The resulting coronal field is then 10G. Two polarisations of driver are used. In the first Alfv\'{e}n waves are 
introduced by driving the bottom boundary out 
of the plane of the simulation and will be called an Alfv\'{e}nic driver. In the second a driver velocity component in the plane of the simulation
is added that 
drives kink waves and thus both Alfv\'{e}n waves and kink waves are present - this is called the mixed mode driver. 
Three options for the spectrum of both drivers have been tested. The first
consists of a low frequency region where the power increases with k and a Kolmogorov region where power drops as $k^{-5/3}$ as in 
equation \ref{driver}.

\begin{equation}
v_{z,x}=A \bigl(\sum \limits_{i=0} \limits^{N_1} \omega_i^1 \sin{(\omega_i t + \phi_i)} + \sum \limits_{i=N_1} 
\limits^{N} \omega_i^{-\frac{5}{6}} \sin{(\omega_i t + \phi_i)}\bigr)
\label{driver}
\end{equation}

\begin{figure}
\includegraphics[scale=0.5]{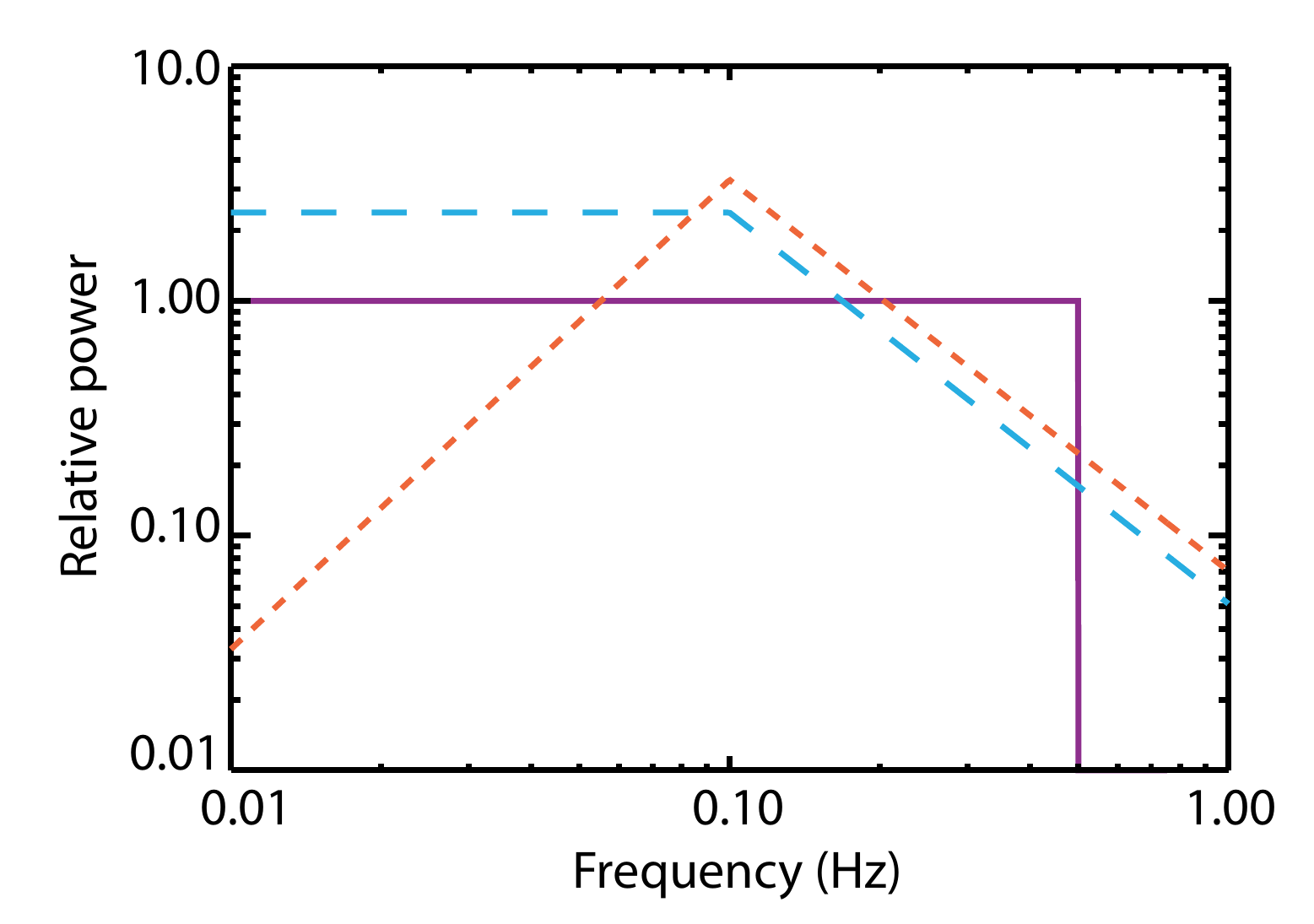}
\caption{Power spectral density of the lower boundary driver used for the simulations. The orange dashed line is from equation \ref{driver}, 
the blue line is equation \ref{driver} above $N_1$ but flat for low frequencies and the solid purple line is a flat spectra with no power-law
dependence. All are normalised to give the same total Poynting flux on the lower boundary.}
\label{fig:driver}
\end{figure}

\change{The velocity component out of the plane of the simulation, $v_z$, is associated with an Alfv\'{e}nic perturbations. 
$v_x$ is the velocity component in the plane of the simulation and is only present in mixed type driving. 
$N$ is the total number of frequencies combined to produce the driver spectrum and $N_1$ is the number of frequencies before 
the maximum in the spectrum. $N$ is a large enough number to ensure that the spectrum is reproduced 
smoothly and that further increase in $N$ doesn't lead to changes in the heating rate of greater than 1\%. $N$ is typically set to 
5,000 in these simulations. $\omega_i$ is the frequency of spectral component $i$ and is logarithmically spaced in the range 0.01Hz - 1 Hz. 
$N_1 = 0.1$Hz for all simulations. $\phi_i$ is the phase for spectral component $i$ and is selected randomly for the $v_z$ component. 
For $v_x$, when present, its phase is the phase of $v_z$ rotated by 90 degrees. The amplitude $A$ is selected for most simulations to give a 
Poynting flux averaged across the whole bottom boundary of 
$2 \times 10^{7} \mathrm{ergs \:cm^{-2} \:s^{-1}}$. A similar driver has previously been used in \citep{Tu2013}. 
For the mixed driver the use of a single amplitude $A$ ensures equal power input to both the Alfv\'{e}nic and kink wave components. Other 
simulations are run with driver Poynting fluxes of $1 \times 10^{8} \mathrm{ergs \:cm^{-2} \:s^{-1}}$ and 
$4 \times 10^{6} \mathrm{ergs \:cm^{-2} \:s^{-1}}$ to evaluate the effect of the driver amplitude on the heating rate.
To assess the robustness of the results to choice of driver some simulations were repeated with a flat driving spectrum up to $N_1$ followed
by a Kolmogorov power law, as in equation \ref{driver}, for higher frequencies. Also tested was 
a flat spectrum with no power-law dependence, see figure \ref{fig:driver}. In results where the different drivers are compared 
the amplitudes are always adjusted so that the total driver Poynting fluxes through the lower boundary are equal.}

\changeb{An isotropic 2D Kolmogorov spectrum will have $E(k) dk \sim v_k^2\ k dk$ in the inertial range. For the purely Alfv\'en wave driver
the $\mathbf{k}$-vector is field aligned and for the simulated flux tube this is 
vertical and thus the spectrum $E(k)\ dk = v_k^2\ dk$ is used as it would be in 1.5D. 
For the mixed mode driver there is a component of the driven wave spectrum
$\mathbf{k}$-vector perpendicular to the field at the boundary. However the wavelength of this component is of the order of the width of
the tube, i.e. 20 km, and this is therefore not in the inertial range for the driver. Thus despite being a 2.5D simulation all results
use the 1.5D $v(k)\sim k^{-5/6}$ driven spectrum which will give an injected energy spectrum of $E(k)\sim k^{-5/3}$.}

\change{The highest frequencies driven into the domain are when the spectra is cut-off at 1Hz. This upper cutoff 
frequency was chosen based on earlier 1D work \citep{Arber2016} where it was shown 
to provide a converged answer. For the atmospheric and magnetic field model used there are a minimum of 3 grid points along a 1Hz Alfv\'{e}n 
wave at the base of the domain where the wavelength is shortest. The wavelength, and thus the number of grid points per wavelength, increase with height.
Note also that there is relatively little power in these high-frequency modes, down by two order of magnitude on the main low-frequency
components.
Testing with a higher upper cutoff frequency, and doubled resolution, shows no change in heating rate or profile. The 
transverse structure of the driver is a Gaussian centered on the bottom boundary flux patch. It is given a 10km width, the same as the flux 
patch.}

\change{The upper boundary of the domain is open. This is implemented using a Riemann characteristic method combined with a damping region.
The damping scales all components of velocity so that on each time-step the velocity calculated by the core solver, $\mathbf{v}$, is replaced by 
$\mathbf{v} \exp(-(y-(l_y-l_{damp}))/(2 l_{damp}))$ 
where $l_y$ is the height of the simulation box and the damping is
only applied above $l_y-l_{damp}$. $l_{damp}$ was set to 500km in the production simulations and was tested for values between 250km and 1000km. 
Varying $l_{damp}$ over this range made $< 0.1$\% different to the heating rate at any height. Testing this boundary with discrete pulses shows 
that less than 0.1\% of the energy incident on the upper boundary returns to the domain. The results presented here are for periodic transverse 
boundary conditions.}

For this initial atmospheric model, magnetic field and boundary driving we solve the compressible, resistive-MHD equations in 2.5D. 
The resistive terms include both contributions from electron-ion plus electron-neutral collisions and the Pedersen resistivity. 
The Pedersen resistivity 
only acts on current perpendicular 
to the magnetic field and results from ion-neutral collisions in the partially ionised chromospheric atmosphere. Neutrals are also required
to get the correct pressure scale height from the C7 model temperature. To correctly handle shocks in these simulations a compatible
shock viscosity \citep{Caramana1998} is used to ensure the correct jump conditions. This also allows measurement of the shock heating which in these
simulations is entirely due to this viscosity. \change{The heating from this shock viscosity is essential to get the entropy jump across the
shock, so this heating is included in the simulation. The model does 
not include thermal conduction or radiative losses and thus throughout the
simulation the atmosphere begins to heat up as we have the heating sources but not the losses. Further simulations are therefore run where 
either the shock heating is simply not included, or a cooling term is included that subtracts a running average over $\tau =160$ seconds 
of the shock heating from the system. Thus the energy equation used in the simulations is,
\begin{equation}
\frac{D \epsilon}{Dt} = -\frac{P}{\rho} \nabla . \mathbf{v} + \frac{H_{\mathrm{visc}}}{\rho} - \frac{H_{\mathrm{cooling}}}{\rho}\label{eq:energy}
\end{equation}
where $\rho$ is the mass density, $P$ is the gas pressure, $\mathbf{v}$ is the fluid velocity, $\epsilon$ is the specific internal 
energy density, $H_{\mathrm{visc}}$ is the viscous (shock) heating and $H_{\mathrm{cooling}}$ is a cooling term given by
\begin{equation}
 H_{\mathrm{cooling}}(\mathbf{r},t)=\frac{1}{\tau}\int_{t-\tau}^t H_{\mathrm{visc}}(\mathbf{r},t')\ dt'
\end{equation} 
This cooling term will tend, on average, to maintain the atmosphere close to its initial profile irrespective of the magnitude of the heating term.
For all simulations $H_{\mathrm{Ohmic}}$, total resistive heating including both electron collisional and Pedersen
resistivities, is calculated for diagnostic purposes but not added to the energy equation. Throughout this paper the only heating term
included in the energy equation is therefore the viscous shock heating.}

\section{Results}

\begin{figure}
\includegraphics[scale=0.5]{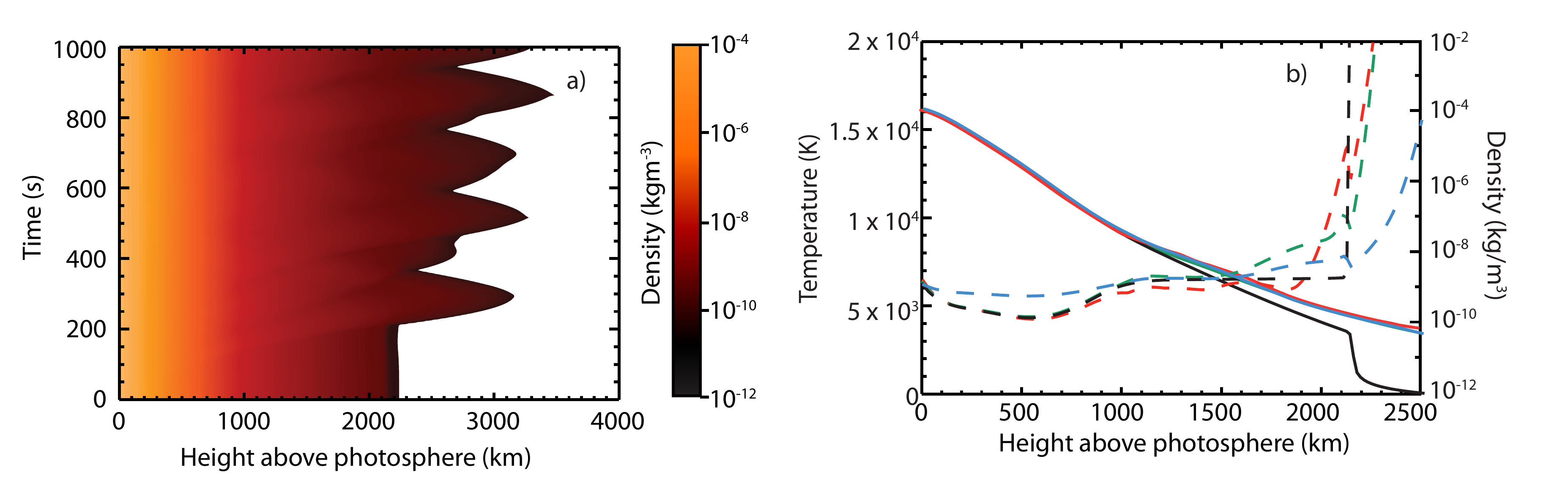}
\caption{Change in equilibrium profiles during the simulations. (a) Time-distance plot of log(density) for densities greater than
$2\times10^{-12}\ \mbox{kg m}^{-3}$ along the line $x=0$. Regions of density below $2\times10^{-12}\ \mbox{kg m}^{-3}$ are shown as white to emphasise the motion of the transition region. (b) Density (solid lines) and temperature (dashed lines) for different 
simulations averaged over the last 100 seconds of the simulation. Black is initial values, blue is the simulation including the 
viscous heating, red is the simulation with the cooling term and green is with no viscous heating in the simulation. 
All results are for the mixed driver with spectrum specified by equation \ref{driver}.}
\label{fig:difference}
\end{figure}

\change{The temperature and density at the start and the end of the simulations, for a variety of combinations of heating terms in equation 
\ref{eq:energy}, are shown in figure \ref{fig:difference} along with a time distance plot of the density. The uplifting of denser material
leads to an increase in the average density for heights 1.2 Mm above the lower boundary. This change is insensitive to the inclusion or
absence of shock heating or the cooling term in equation \ref{eq:energy}. The temperature profile does change depending on the terms
included in equation \ref{eq:energy} and so the simulations below are repeated for all three cases, i.e. no heating, shock heating and shock heating
plus cooling. The primary difference between these is that the simulations which include shock heating, but not the cooling term, 
lead to a less steep temperature profile through the transition region as the atmosphere continues to heat and rise. In this case the 
majority of the additional heating goes into an increase in the gravitational potential energy.}

\changeb{Figure \ref{fig:difference} b) shows that the force balance in the upper chromosphere and transition region is changed by the action of the waves. Temperature and pressure profiles change differently, so considering only gravity and static pressure forces there is a net downwards force. Despite this figure \ref{fig:difference} a) shows that the atmosphere is still in a dynamic equilibrium state, oscillating vertically but not systematically moving up or down. The additional upwards pressure force is provided by the shock ram pressure $0.5 \rho v^2$ which in the upper chromosphere is 2.5 times the static pressure, consistent with the observed shock Mach numbers.}

\begin{figure*}
\includegraphics[scale=0.5]{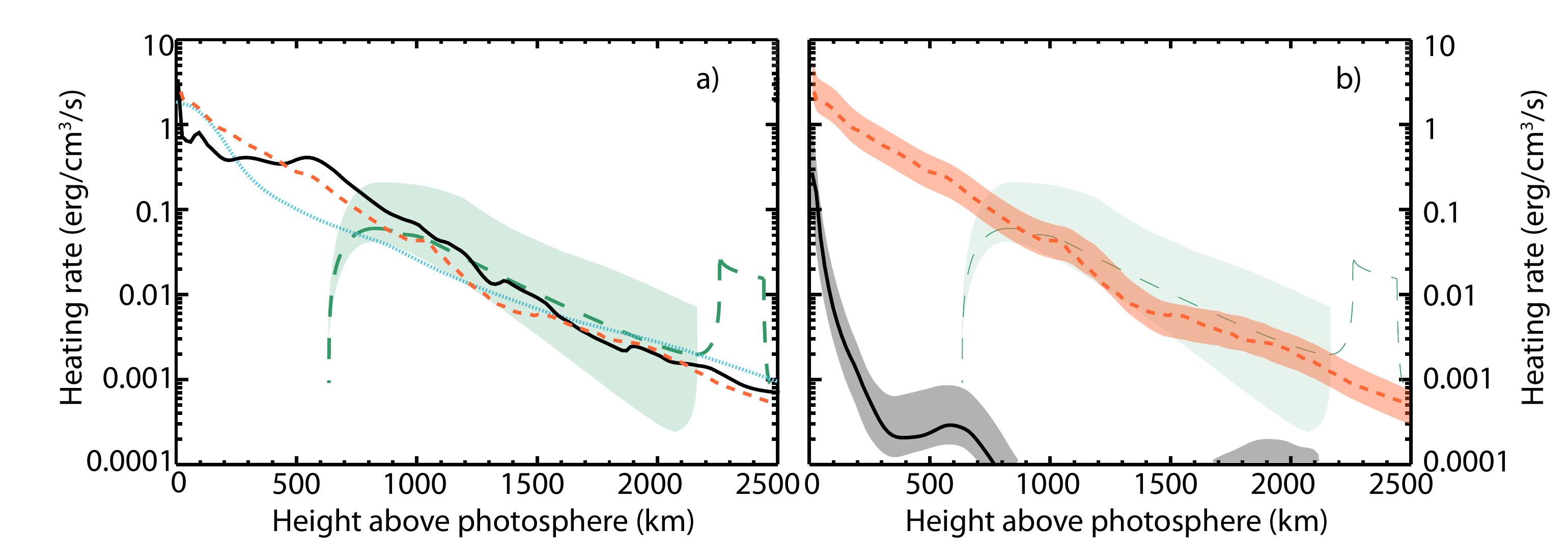}
\caption{Heating rates as a function of height averaged across the simulation domain. \change{On panel a)} the solid black line is 
viscous heating from simulations with mixed mode driving. The blue dotted line is the 
heating from a purely Alfv\'{e}nic driver. \change{The orange short 
dashed line is from a simulation with a mixed mode driver including the cooling term in the energy equation. 
The solid black, blue dotted and orange dashed line are all from simulations
which included the viscous heating in the energy equation and used the spectrum specified by equation \ref{driver}.} The green dashed line is the 
estimate of local cooling due to chromospheric radiation from \citet{Avrett1981} Model C for the quiet chromosphere. The green shaded area is bounded 
by the Avrett Model A (dark network region) and Avrett Model F (very bright network element). \change{On panel b) the orange short 
dashed line, green dashed line and green area are the same as in panel a). The black solid line is the resistive heating from the 
same simulation, and the grey and orange areas are bounded by simulations where the Poynting flux through the boundary is either increased 
or decreased by a factor of 5.}
}
\label{fig:heating}
\end{figure*}

The \change{viscous} heating rate averaged across the flux structure is shown in figure 
\ref{fig:heating} a) as a black line for a mixed mode driver and a blue dotted line for the Alfv\'{e}nic driver. 
\change{The orange dashed line shows the effect of including the cooling term in the mixed driver simulation.} 
\change{Simulations for the mixed driver but without either the cooling term or
viscous heating included in the energy equation produce a heating profile indistinguishable from the orange dashed line in
figure \ref{fig:heating}(a) and is therefore not shown.}
Comparing with observed heating rates from \citet{Avrett1981} (green dashed line and grey shaded area), the simulated heating rates are a match from 700km above the photosphere to 2100km. 
\change{The driver amplitude, and hence Poynting flux, are only loosely constrained by observations to be close to 
$2 \times 10^{7} \mathrm{ergs \:cm^{-2} \:s^{-1}}$ so the effect of changing this is shown in 
figure \ref{fig:heating} panel b). 
The orange dashed line is the same as in panel a) and the orange shaded region is the range of heating rates obtained by 
increasing or decreasing the driver Poynting flux by a factor of 5. The black line is the resistive heating rate, and the gray 
shaded region around it is the change in the resistive heating by changing the driver Poynting flux. The heating rate from the 
simulations are in quantitative and qualitative agreement with the observationally determined heating from \citet{Avrett1981} 
between the heights of 700km and 2100km. In \citet{Avrett1981} it is explicitly stated that "The net radiative cooling rate $\phi(h)$ 
throughout the chromospheric portion of each model is a direct measure of the non-radiative heating required to produce the 
chromospheric temperature increase.", and that the apparent temperature plateaus in the transition region 
in Avrett's model that are needed to 
reproduce the Lyman-$\alpha$ spectrum make interpretation of the radiative loss functions above 2 Mm difficult. The use of 
radiation from chromospheric lines in 
\citet{Avrett1981} also mean this method cannot match heating requirements in the photosphere. Thus one would only expect the 
shock heating
calculated here from simulations to match the results from 
\citet{Avrett1981} in the range shown shaded in figure \ref{fig:heating} between 700-2100 km.} 

\begin{figure*}
\includegraphics[scale=0.5]{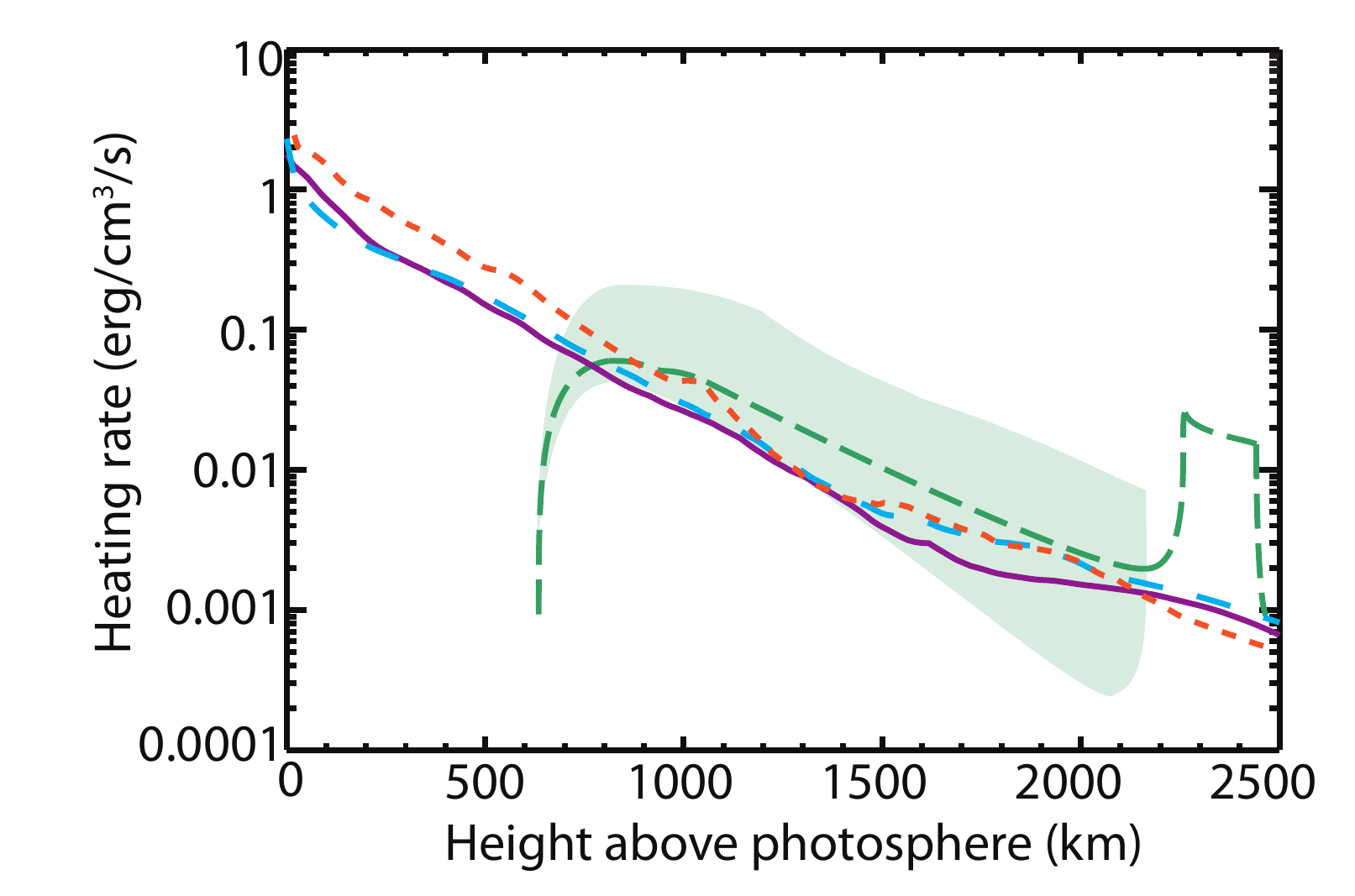}
\caption{\change{Heating rates as a function of height averaged across the simulation domain for various driving spectra.
All results are for the mixed mode driver with shock heating and the cooling term in the energy equation. The orange
dashed line 
is for a spectrum matching equation \ref{driver} and is the same as the orange line in figure \ref{fig:heating}. 
The blue line is equation \ref{driver} above $N_1$ but flat for low frequencies and the solid purple line is a 
flat spectra with no power-law dependence. Colours match those use in figure \ref{fig:driver}}
}
\label{fig:spec-heat}
\end{figure*}

\change{To assess the sensitivity of the heating profiles to the choice of spectra the results for the mixed driver with viscous
heating and cooling terms included in the energy equation were repeated for the three spectra shown in figure \ref{fig:driver}. 
These results are shown in figure \ref{fig:spec-heat} confirming that the precise functional form of the driving spectrum
only weakly affects the heating profile. Across all driver spectra roughly 50\% of the boundary driven Poynting flux
emerges into the corona.}

\begin{figure*}
\includegraphics[scale=0.5]{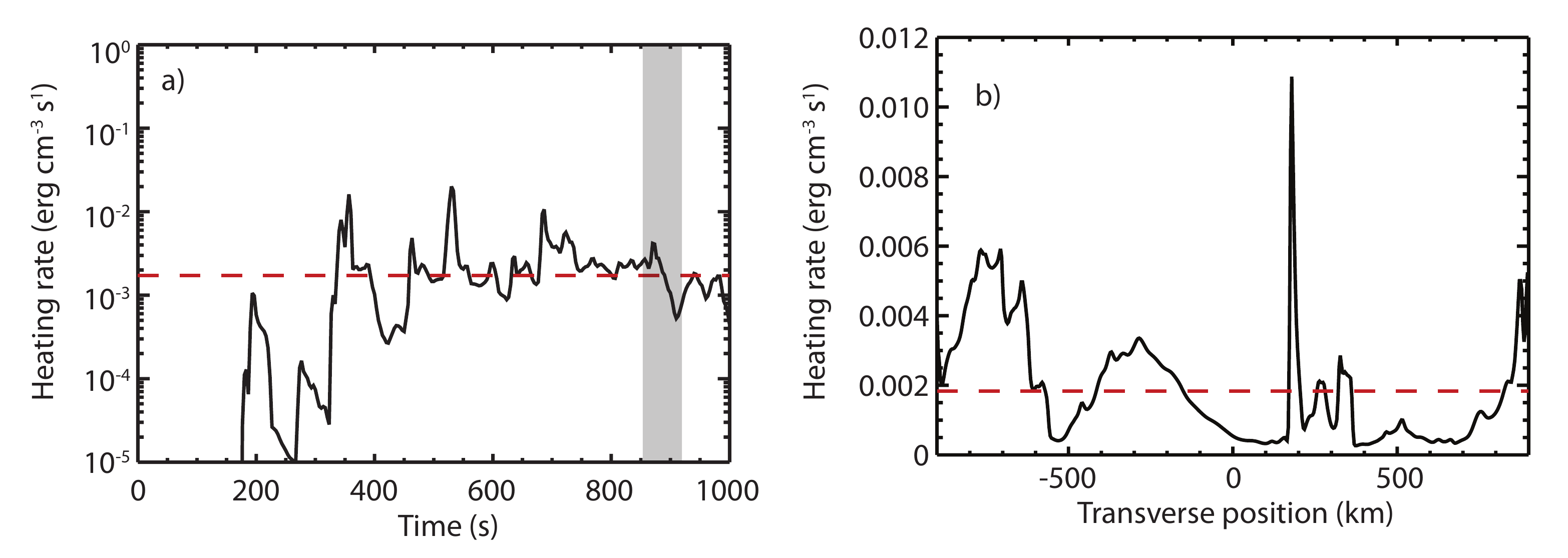}
\caption{\change{Panel a) shows the time history of the instantaneous heating rate at a point 100km below the initial height of the 
transition region at the centre of the flux structure for a simulation of a mixed mode driver with included cooling term. The red 
dashed line is the average heating rate for that height as shown on figure \ref{fig:heating}. Panel b) shows the transverse 
structure of the heating at the same height averaged over the time shown in the grey box in panel a. The red dashed line is 
again the average heating rate.}}
\label{fig:transverse}
\end{figure*}

\change{
The averaged heating rates in figure \ref{fig:heating} miss the variability of the shock heating. An example of the 
variability 100km below the transition region is shown in figure \ref{fig:transverse} 
panel (a). Here the height of the initial transition region is defined as the height at which the initial equilibrium
mass density $\rho=10^{-11}\mbox{kg m}^{-3}$, i.e. a height of 2162 km for this simulation domain. An example of the 
transverse structure of the heating rate averaged over the grey box in panel (a) is shown in panel (b), demonstrating that transverse 
structure is present in the simulation.}

\begin{figure}
\includegraphics[scale=0.5]{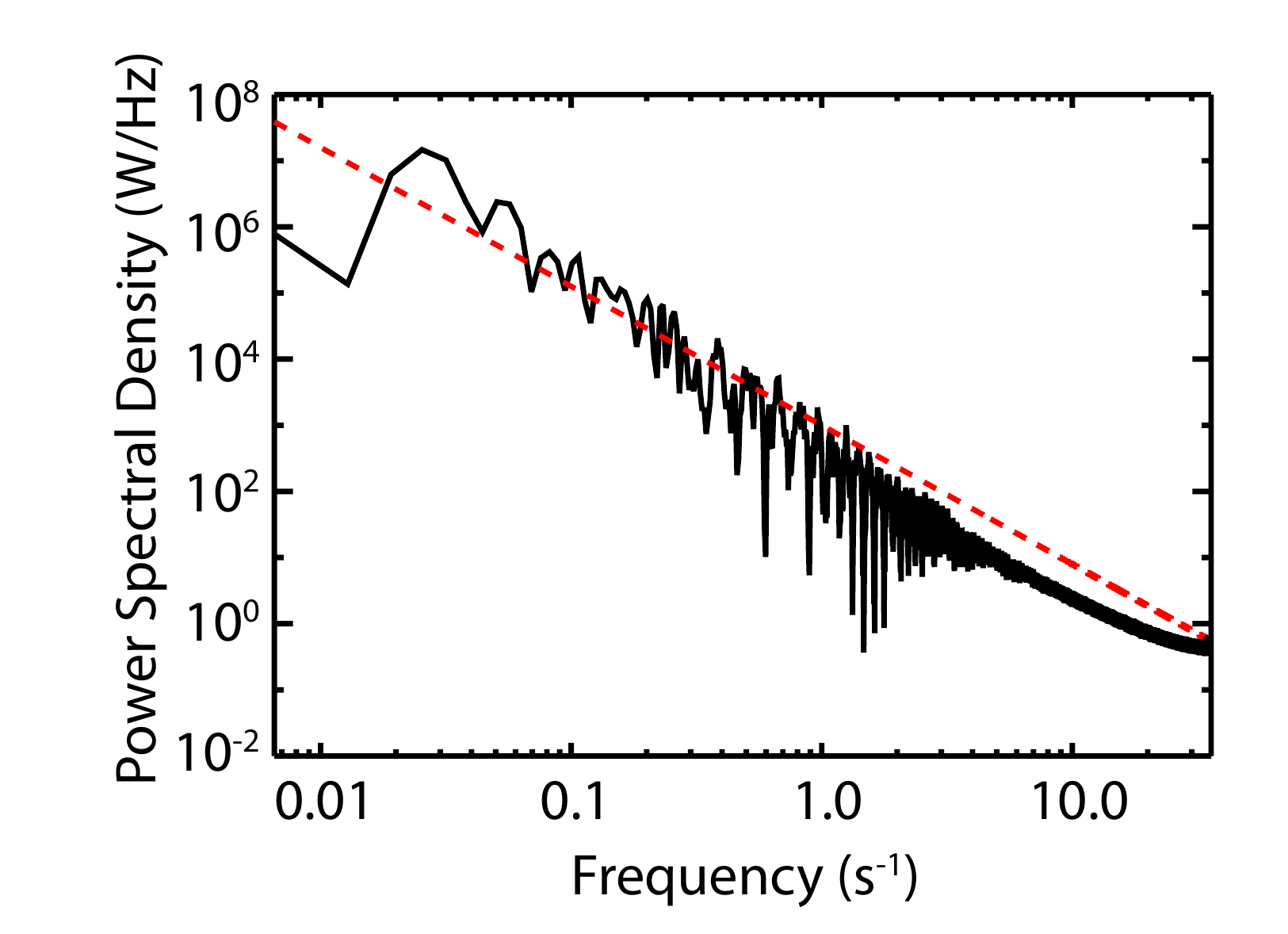}
\caption{Power spectral density measured at a point 100km below the original location of the transition region. The red 
dashed line is a power law with a power of -2.}
\label{fig:upper_power}
\end{figure}

\begin{figure}
\includegraphics[scale=0.5]{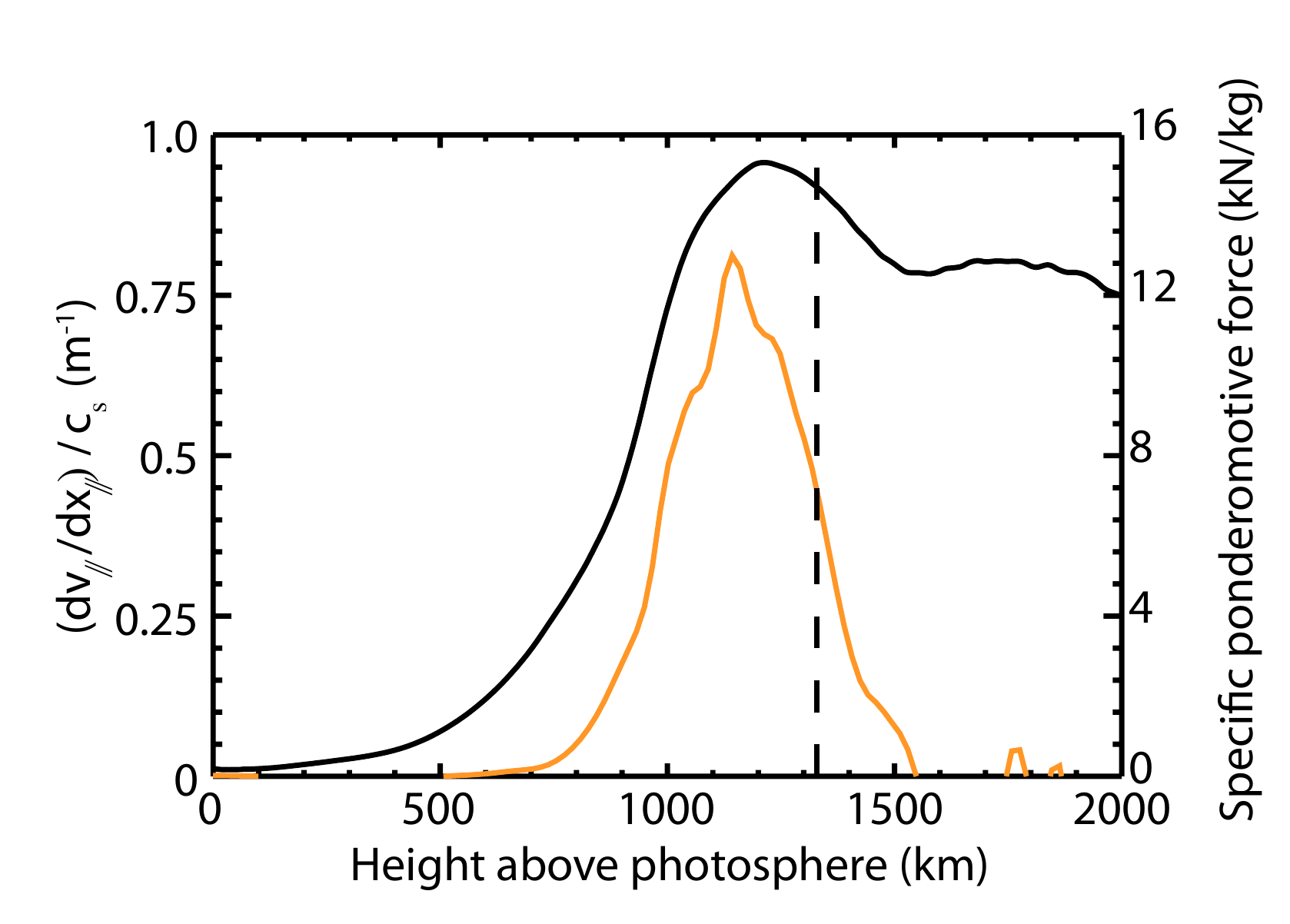}
\caption{The specific ponderomotive force (black line) and parallel compressive component of $\nabla.\mathbf{v}$ normalised to 
the  local sound speed (orange line) for the mixed mode driver. Both are averaged across the flux tube at fixed height and over
the longest period of the boundary driver. The vertical line is the location of the $\beta=1$ surface.}
\label{fig:force}
\end{figure}

The origin of the heating is from shock dissipation of acoustic modes propagating along field lines. \change{This is shown in 
figure \ref{fig:upper_power} by the presence of the -2 power law which is characteristic of shock dominated time series.
Note that while the shortest wavelength which can be resolved at $y=0$ restricts the driver to frequencies below 1 Hz the
wavelength expansion with height means that the spectrum in figure \ref{fig:upper_power} extends beyond 1 Hz.} 
\change{The heating rates recorded in these simulations are purely those due to shock viscosity and resistivity. Heating due 
to resistivity is shown to be small and can be ignored. Shock viscosity goes to zero in smooth regions of the 
solution, and the convergence of the numerical results gives confidence that the heating is due to 
discontinuities, i.e. shocks. It is possible that phase mixing, that cascades down to 
arbitrarily small scales in the absence of bulk viscosity, may contribute to some of the heating. However 
the fact that the evolved velocity spectrum is that of a shock 
dominated time series strongly suggests that the heating is dominated by shocks. In addition the heating is strong in the centre
of the flux tube, where phase-mixing would be absent.} There is no acoustic component in the driver so 
these waves are generated from the boundary driven MHD waves. Figure \ref{fig:force} shows both the specific ponderomotive force 
and parallel compressive component of $\nabla.\mathbf{v}$ for the mixed mode driver averaged across the simulation domain and 
the longest period in the driver. The specific 
ponderomotive force is defined as \change{$(-1/\mu_0 \rho)\nabla B_\perp^2$ where $B_\perp$ is the magnetic field perturbation 
perpendicular to the background magnetic field.}

\begin{figure}
\includegraphics[scale=0.5]{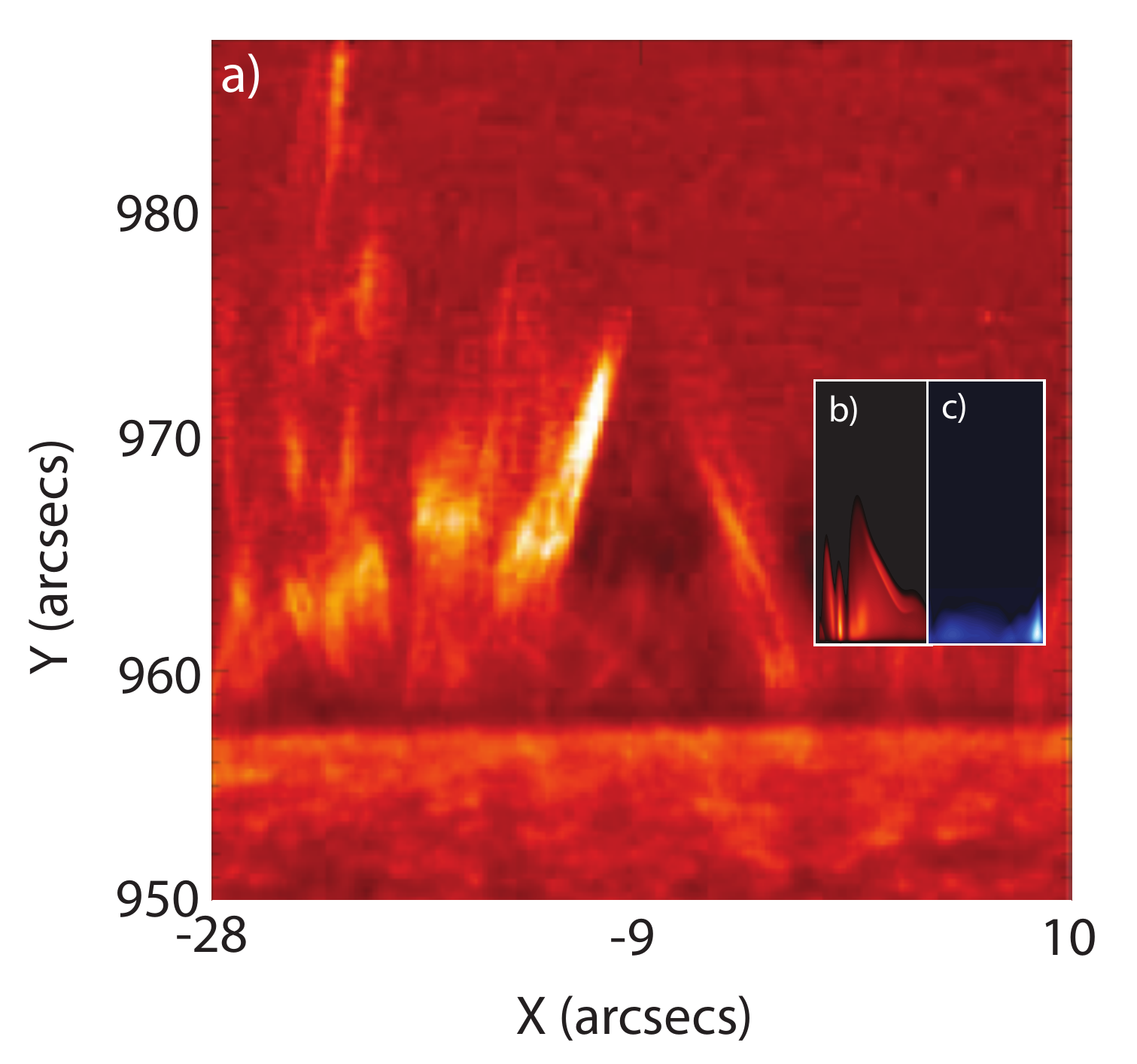}
\caption{Comparison of observed and simulated spicules. The observation is from an Hinode SOT Ca II H image showing 
spicules taken on April 29th at 
0224UT from \citet{Tsiropoula2012}. The insert plots are the relative change in density from the initial conditions 
of the coronal part of a simulation 
at the same height and scale. The red coloured insert (b) is for the mixed mode driver and the blue insert (c) for the purely
Alfv\'{e}nic driver.}
\label{fig:spicule}
\end{figure}

\begin{figure*}
\includegraphics[scale=0.5]{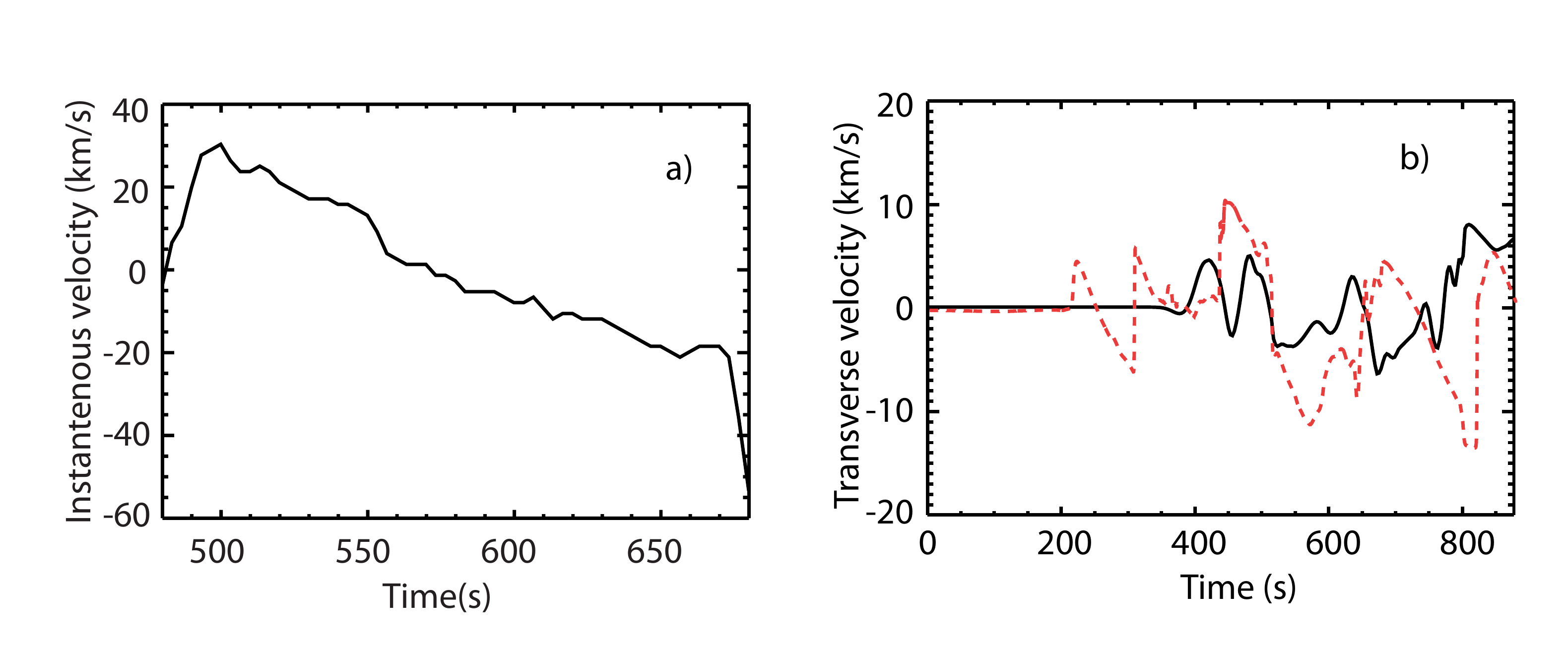}
\caption{\change{a) Instantaneous vertical velocity of a spicule measured as the speed of mass density 
$\rho=10^{-11}\mbox{kg m}^{-3}$ b) Transverse velocity 
(out of the simulation plane) measured at transition region height from the centre of the flux tube. The solid line is for the Alfv\'{e}nic driver
and the red dashed line the mixed driver.}}
\label{fig:spicule_osc}
\end{figure*}

The formation of shocks causes a lifting of dense material from the lower atmosphere (figure \ref{fig:spicule}), and for the mixed mode drivers, 
internal reflections in the flux tube cause transverse structuring of this dense material. If a purely Alfv\'enic driver is used then
the transverse structuring in the shock driven, uplifted material is reduced. Taking the relative density change of the plasma 
in the simulation and overplotting it on an example observational picture of spicules produces a strong visual similarity, so these 
density structures can be preliminarily described as spicules. These spicules are carried up to a height of $\sim 2500-4000$ km above the photosphere.
\change{The rise speed of spicules is not the local fluid speed as these are propagating shocks. The rise speed is therefore calculated
by tracking the location of the reference transition region mass density $\rho=10^{-11}\mbox{kg m}^{-3}$ for rising spicules. The
mean rise speed across all spicules was ${\sim} 12 \mathrm{\:km\:s^{-1}}$. A representative spicule is shown in figure \ref{fig:spicule_osc} (a)
for which the peak rise speed is $30 \mathrm{\:km\:s^{-1}}$ and the average rise speed $15 \mathrm{\:km\:s^{-1}}$. The range of maximum rise
speeds across all spicules is $24-32 \mathrm{\:km\:s^{-1}}$ and the range of average rise speeds is $10-15 \mathrm{\:km\:s^{-1}}$
}
Both the height and the rise speed are comparable with observations (1000-10000km above the photosphere and 
15-65km/s including both Type I and Type II spicules. These results best match Type I spicules.) 
(\citet{Beckers1968}, \citet{Pasachoff2009}, \citet{Pereira2012}). Recording the out-of-plane velocity at the transition 
region (figure \ref{fig:spicule_osc} b),
transverse oscillations are measured with an r.m.s. speed of $\sim 9 \mathrm{\:km\:s^{-1}}$ for the mixed driver
again consistent with observations (\citet{Depontieu2007,Zhang2012}).

\changeb{The results are qualitatively in agreement with the results of \citet{Matsumoto2014} who briefly discuss 
heating of the chromosphere in their section 6.1. They find that heating below the transition region is dominated 
by shock heating and identify both fast and slow shocks as being important. The overall heating rates in
\citet{Matsumoto2014} are comparable to those presented here but due to the the low resolution in the chromosphere
used in those simulations \citet{Matsumoto2014} point out that their code produces strong numerical 
heating. The results presented in this manuscript are fully converged.}

\section{Conclusions}
This letter presents the results from 2.5D simulations of MHD wave driven heating of the chromosphere which also show 
spicule launching. They show that by choosing plausible values for the driver amplitude and spectrum, magnetic field and atmosphere model 
robust heating rates comparable to observations are obtained. Self-consistently with this, spicule like blobs of enhanced density are launched 
into the corona with average rise speeds of ${\sim} 12 \mathrm{\:km\:s^{-1}}$, peak rise speeds in the range 
$24-32 \mathrm{\:km\:s^{-1}}$. These spicules have transverse oscillations with r.m.s. velocities of 
${\sim} 9 \mathrm{\:km\:s^{-1}}$. All of these numbers are consistent with observations. If a mixed kink and Alfv\'{e}n type driver is specified 
then realistic transverse structuring of the driven spicules is also observed. From the simulations presented here we conclude:
\begin{itemize}
   \item For drivers which are either Alfv\'{e}n waves or mixed Alfv\'{e}n and kink waves, provided the Poynting flux is kept near
   ${\sim} 2 \times 10^7 \mathrm{erg\:cm^{-2}\:s^{-1}}$, these waves can heat the upper chromosphere.
   \item This heating is from shock dissipation of slow modes generated by the ponderomotive force.
   \item The heating is insensitive to the upper cut-off in driver spectrum or profile across the three spectra tested.
   \item Compressibility is essential for modelling the chromosphere and an Alfv\'{e}n wave driver does not lead to a turbulent cascade
   to dissipation scales but instead loses energy by coupling to slow modes.
   \item If a mixed mode driver is specified internal reflection within the flux tube generates transverse structure higher up 
   similar to that observed in spicules. 
   \item The rise of dense material has a velocity consistent with Type-I spicules.
   \item The transverse oscillations have peak and RMS velocities comparable to observations. 
\end{itemize}

\change{These simulations are limited to 2.5D. The fastest Alfv\'{e}n wave cascade in Alfv\'{e}nic turbulence requires
wave-vector matching in 3D. It is therefore possible that in 3D a turbulent cascade terminated by resistive damping will be a more efficient 
heating mechanism. However, given that shock heating is orders of magnitude more important than resistive in 2D, and 3D effects 
will not turn off shock heating, shock heating is still likely to remain an important heating mechanism in 3D. }

The model used in this paper does not attempt to accurately describe the non-local transport of energy through radiation or 
non-LTE physics. These are 
surely important for predicting observational signatures of chromospheric spectral lines. 
Despite this the heating rates and spicule properties have converged before the simulation ends and the coupling of
MHD waves to slow modes via the ponderomotive force depends only on the magnetic field and local mass density. 
\change{This mechanism is robust in that the ponderomotive force depends on only the gradient of the MHD wave magnetic field energy
which in turn acts on the local chromospheric mass density. This occurs low in the atmosphere where the ion-neutral coupling
is strong and hence all that matters is the MHD wave energy, its spectrum, and the local mass density. Once these waves are
generated they will shock due to density stratification in the upper atmosphere - thereby heating the chromosphere.}
As such, it seems likely that the results presented here would be reproduced in a  full radiation hydrodynamic, non-LTE simulation
of the same shock resolution.  
These results show that both the qualitative and quantitative properties of chromospheric heating and Type-I spicules can be 
produced from the same underlying 
physical process: the ponderomotive formation of shocks from transverse photospheric motion.

\section*{Acknowledgements}
This work used the Darwin Data Analytic system at the University of Cambridge, operated by the University of Cambridge High Performance Computing Service on behalf of the STFC DiRAC HPC Facility (www.dirac.ac.uk). This equipment was funded by a BIS National E-infrastructure capital grant (ST/K001590/1), STFC capital grants ST/H008861/1 and ST/H00887X/1, and DiRAC Operations grant ST/K00333X/1. DiRAC is part of the National E-Infrastructure.\\

 This work used the DiRAC Data Centric system at Durham University, operated by the Institute for Computational Cosmology on behalf of the STFC DiRAC HPC Facility (www.dirac.ac.uk). This equipment was funded by a BIS National E-infrastructure capital grant ST/K00042X/1, STFC capital grant ST/K00087X/1, DiRAC Operations grant ST/K003267/1 and Durham University. DiRAC is part of the National E-Infrastructure. \\
 
\bibliographystyle{apj}
\providecommand{\noopsort}[1]{}\providecommand{\singleletter}[1]{#1}%

\end{document}